\documentclass{article}

\usepackage{arxiv}

\usepackage{cite}
\usepackage{amsmath,amssymb,amsfonts}
\usepackage{algorithm}
\usepackage{algorithmic}
\usepackage{graphicx}
\usepackage{textcomp}
\usepackage[table]{xcolor}
\usepackage{booktabs}
\usepackage{multirow}
\usepackage{array}
\usepackage{url}
\usepackage{tikz}
\usetikzlibrary{arrows.meta,positioning,shapes.geometric,calc,backgrounds,fit}
\usepackage{pgfplots}
\pgfplotsset{compat=1.18}
\usepackage[hidelinks]{hyperref}

\def\BibTeX{{\rm B\kern-.05em{\sc i\kern-.025em b}\kern-.08em
    T\kern-.1667em\lower.7ex\hbox{E}\kern-.125emX}}

\newcolumntype{L}[1]{>{\raggedright\arraybackslash}p{#1}}

\definecolor{tableheader}{HTML}{1C355E}
\definecolor{rowtint}{HTML}{EEF2F8}
\definecolor{boxtint}{HTML}{E8EEF7}
\definecolor{gatetint}{HTML}{FBE9D6}

\newtheorem{definition}{Definition}

\newcommand{\spec}{\ensuremath{\mathcal{S}}}
\newcommand{\impl}{\ensuremath{I}}
\newcommand{\gen}{\ensuremath{G}}
\newcommand{\val}{\ensuremath{V}}
\newcommand{\accept}{\ensuremath{\mathcal{A}}}

\title{Specification-Driven Development as the Foundation of AI-Native Enterprise Software Engineering}

\author{
 Mamdouh Alenezi \\
  Saudi Data and Artificial Intelligence (SDAIA)\\
  Riyadh, Saudi Arabia
}

\begin{document}
\maketitle
\begin{abstract}
Large language models (LLMs) and agentic AI systems are shifting software engineering from manual coding toward intent specification, architecture, verification, and governance. Two contrasting paradigms have emerged: vibe coding, a conversational, intuition-driven approach where AI-generated artifacts are accepted primarily by observed behavior, and Specification-Driven Development (SDD), which treats structured, machine-readable specifications as the authoritative source of truth from which implementations are generated and verified. This article makes three contributions. First, drawing on a systematically verified corpus of peer-reviewed and preprint literature, it identifies recurring failure modes of ungoverned conversational generation, including the productivity-reliability paradox, architectural erosion from limited model context, increased security exposure, and accumulating technical debt. Second, it introduces the Specification Governance Reference Model (SGRM), a tool-independent framework that defines four-component specification contracts, constrains stochastic generation within deterministic validation, formalizes three specification rigor levels, and integrates generation, verification, and governance into a closed-loop architecture. Third, it evaluates SGRM against ISO/IEC 25010, mapping quality characteristics to governance mechanisms and supporting empirical evidence, including enterprise studies reporting a 73\% reduction in security defects under constitutional specification constraints and a 50\% reduction in time-to-market through specification-governed agentic delivery. The analysis argues that while vibe coding is valuable for ideation and prototyping, enterprise-grade software requires specification governance to transform probabilistic AI generation into deterministic, auditable engineering. Boundary conditions, threats to validity, and future research directions are discussed.
\end{abstract}


% ================= INTRODUCTION =================
\section{Introduction}
\label{sec:introduction}

The traditional software development lifecycle treated source code as the definitive blueprint and ultimate source of truth for an application: requirements were translated by human hands into syntax, and that syntax constituted the durable, version-controlled asset around which testing, review, and maintenance were organized. The maturation of large language models (LLMs) and multi-agent systems has disrupted this hierarchy by separating cognitive architectural design from syntactic implementation. In the emerging model, developers declare intent and interface constraints while autonomous systems synthesize code as a downstream---and, in the limit, transient---byproduct \cite{alenezi2026rise,piskala2026spec}. Researchers have framed this transition as the arrival of \emph{AI-native software engineering}, sometimes labeled Software Engineering~3.0, in which AI acts as a collaborative agent integrated into every phase of the lifecycle rather than as an auxiliary autocomplete tool \cite{hassan2024towards,alenezi2026rise}. In this regime, code transitions from a scarce, laboriously crafted artifact to an abundant, disposable commodity, and the discipline must reorganize around strategic orchestration of multi-agent workflows, rigorous verification of AI-generated outputs, and structured human--AI collaboration \cite{alenezi2026rethinking}.

This shift is not automatically benign. The speed of AI-assisted generation can compromise the structural integrity of the resulting systems if it is not governed by methodological frameworks designed for the new division of labor \cite{ozkaya2023application,fawzy2026vibe}. Practice has polarized around two archetypes. The first, colloquially termed \emph{vibe coding} after a widely circulated post by Karpathy \cite{karpathy2025vibe}, is a conversational, intuition-driven mode of AI-assisted development in which software emerges through iterative natural-language dialogue, often without systematic review of the generated artifacts \cite{meske2025vibe,fawzy2026vibe}. The second, \emph{Specification-Driven Development} (SDD), inverts the traditional workflow by elevating structured, machine-readable specifications to the position of authoritative source of truth, from which implementations are generated, regenerated, and verified \cite{piskala2026spec}.

The empirical literature on these two practices has grown rapidly but remains fragmented across venues, publication types, and levels of maturity: controlled experiments on productivity \cite{peng2023impact,borg2026echoes}, grey-literature studies of practitioner behavior \cite{fawzy2026vibe}, security benchmarks and user studies \cite{pearce2022asleep,perry2023users,sandoval2023lost,fu2025security}, methodological treatments of SDD \cite{piskala2026spec,marri2026constitutional}, and enterprise case studies \cite{vilasboas2026one}. What is missing is a unifying, formally articulated account of \emph{why} specification governance addresses the documented failure modes of ungoverned generation, and a systematic assessment of that account against the quality-attribute demands that define enterprise-grade software.

This article addresses that gap. It is guided by three research questions:

\begin{itemize}
\item \textbf{RQ1.} What failure modes of ungoverned, conversational AI code generation does the verifiable empirical evidence identify with respect to enterprise quality attributes?
\item \textbf{RQ2.} Through which formal mechanisms does specification governance address these failure modes, and how can those mechanisms be consolidated into a tool-independent reference model?
\item \textbf{RQ3.} To what extent does the published empirical evidence support specification-driven development as the foundation of AI-native enterprise software engineering, and what are its boundary conditions?
\end{itemize}

In answering these questions, the article makes four contributions:

\begin{enumerate}
\item \emph{An evidence-based failure-mode analysis} (Section~\ref{sec:failures}) that consolidates the verified empirical record on ungoverned AI-assisted development into four recurring failure modes---the productivity--reliability paradox, architectural erosion, security exposure, and technical-debt accumulation---each traced to its underlying causal mechanism.
\item \emph{The Specification Governance Reference Model (SGRM)} (Section~\ref{sec:sgrm}), a formal, tool-independent reference model for AI-native development that (i)~defines specifications as four-component contracts over functional obligations, quality constraints, constitutional constraints, and architectural structure; (ii)~positions the stochastic generator inside a deterministic validation boundary; (iii)~formalizes the three tiers of specification rigor identified in the practitioner literature \cite{piskala2026spec}; and (iv)~organizes these elements into a layered architecture with a closed-loop regeneration algorithm and six falsifiable design propositions.
\item \emph{An analytical evaluation against ISO/IEC~25010} (Section~\ref{sec:evaluation}) that maps each product quality characteristic of the standard \cite{iso25010} to the documented risk it faces under ungoverned generation, the SGRM mechanism that addresses it, and the published evidence supporting that mechanism, explicitly including counter-evidence and unreplicated results.
\item \emph{A critical discussion and research agenda} (Section~\ref{sec:discussion}) that identifies the boundary conditions under which SDD is and is not warranted, the implications for the competency model of the enterprise engineer, and the open empirical questions on which the argument's weakest links depend.
\end{enumerate}

Methodologically, the article follows a design-science structure \cite{hevner2004design,wieringa2014design}: the reference model is the designed artifact, constructed from a systematically verified evidence corpus and evaluated by informed argument against an external quality standard using only published empirical results (Section~\ref{sec:method}). The position defended is deliberately conditional rather than absolutist: vibe coding is a legitimate and valuable instrument for ideation, prototyping, learning, and accessibility \cite{fawzy2026vibe,meske2025vibe}, and SDD carries authoring and maintenance costs that simpler contexts do not repay \cite{piskala2026spec}. The claim is that \emph{where} enterprise-grade quality attributes---testability, scalability, security, maintainability, traceability, and regulatory compliance---are required, specification governance is the mechanism by which they are obtained from stochastic generators.

The remainder of the article is organized as follows. Section~\ref{sec:background} reviews the conceptual foundations of vibe coding and SDD and their historical lineages. Section~\ref{sec:method} describes the research method. Section~\ref{sec:failures} answers RQ1, Section~\ref{sec:sgrm} answers RQ2, and Section~\ref{sec:evaluation} answers RQ3. Section~\ref{sec:discussion} discusses implications and open questions, Section~\ref{sec:threats} examines threats to validity, and Section~\ref{sec:conclusion} concludes.

% ================= BACKGROUND =================
\section{Background and Related Work}
\label{sec:background}

\subsection{The AI-Native Paradigm Shift}
\label{sec:ainative}

The intersection of LLMs and software engineering has produced what a growing body of literature characterizes as a structural transformation of the discipline, reshaping development processes, required competencies, professional roles, and educational outcomes \cite{alenezi2026rise,hou2024large,fan2023large}. Systematic reviews document substantial gains in implementation-phase throughput across code generation, completion, repair, testing, and documentation tasks \cite{hou2024large,fan2023large}, while vision papers argue that the next phase of the discipline---AI-native software engineering---will integrate trustworthy and efficient AI as a first-class collaborator across the lifecycle \cite{hassan2024towards}. The redistribution of cognitive labor is a consistent theme: engineers migrate from syntax production toward formal requirement specification, systematic verification, and system orchestration \cite{alenezi2026rise,alenezi2026rethinking,meske2025vibe}. It is precisely this redistribution that gives the specification a new centrality: when implementation is delegated, the artifact that expresses \emph{what is to be implemented} becomes the primary locus of engineering judgment.

\subsection{Vibe Coding: Definition and Conceptual Foundations}
\label{sec:vibe}

The term \emph{vibe coding} was coined informally by Karpathy in February 2025, describing a free-form conversational style of producing software with LLMs in which the developer ``gives in to the vibes'' and largely ceases to inspect the generated code \cite{karpathy2025vibe}. The practice rests on the demonstrated capacity of neural models to synthesize code from natural-language intent: Codex solved a substantial fraction of programming problems from docstring descriptions alone \cite{chen2021evaluating}, and AlphaCode achieved approximately median-human performance in competitive programming \cite{li2022competition}, validating natural language as a viable programming interface.

Academic formalization followed quickly. Meske et al.\ define vibe coding as ``a software development paradigm where humans and generative AI engage in collaborative flow to co-create software artifacts through natural language dialogue,'' and analyze it as a \emph{reconfiguration of intent mediation}: across nine decades of computing, the mediation of developer intent has evolved from manual hardware manipulation through structured languages to the present shift from deterministic instruction to probabilistic inference \cite{meske2025vibe}. Fawzy et al.\ characterize the practice operationally: coders---including novices and non-developers---rely on AI code generation through intuition and trial-and-error, without necessarily understanding the underlying code or conducting rigorous review and testing \cite{fawzy2026vibe}. Two properties explain both the appeal and the risk of the paradigm. First, it redistributes effort from syntax production toward intent formulation and output evaluation, with acceptance decided by immediate behavioral observation---the ``vibe'' of the current execution---rather than by verification against a pre-defined blueprint \cite{meske2025vibe,fawzy2026vibe}. Second, it democratizes software creation, dramatically lowering the barrier to exploratory programming for non-developers \cite{fawzy2026vibe}. These properties make vibe coding a genuine socio-technical innovation for ideation, rapid prototyping, learning, and experimentation; whether they extend to enterprise-grade systems is the question this article addresses.

\subsection{Specification-Driven Development}
\label{sec:sddbg}

Specification-Driven Development extends classical software engineering by treating structured specifications, rather than source code, as the authoritative source of truth. Piskala, in the most comprehensive treatment to date, defines SDD as an inversion of the traditional workflow: specifications become the primary artifact, and code is generated or verified as a secondary, derived artifact \cite{piskala2026spec}. Requirements, architecture, acceptance criteria, API contracts, business rules, and quality constraints are expressed as machine-readable specifications that guide autonomous generation; instead of repeatedly prompting an LLM with informal requests, developers continuously refine executable specifications from which implementations can be regenerated on demand. Piskala formalizes the methodology along two axes: a maturity model of \emph{specification rigor} comprising three tiers (spec-first, spec-anchored, and spec-as-source, formalized in Section~\ref{sec:tiers}), and a four-phase workflow---\emph{Specify, Plan, Implement, Validate}---that ties intent to implementation through automated checks at each transition \cite{piskala2026spec}. The framework maps onto real tooling: behavior-driven development frameworks, API-contract ecosystems, and AI-assisted toolkits such as GitHub Spec Kit, with case studies in API-first microservices, enterprise feature development, and model-based embedded software \cite{piskala2026spec}. Marri extends the methodology with \emph{Constitutional Spec-Driven Development} (CSDD), in which a versioned, machine-readable ``constitution'' encodes non-negotiable security constraints derived from CWE weakness classes and regulatory frameworks and governs all subsequent generation \cite{marri2026constitutional}.

Methodologically, SDD adapts the foundational logic of test-driven development by positioning the governing artifact further upstream: the specification serves as the comprehensive source from which both executable code and its verification suites are synthesized \cite{piskala2026spec,vilasboas2026one}. By elevating the primary artifact from code to contract, SDD establishes a deterministic boundary that constrains the stochastic output of generative models: the model may propose any implementation, but only implementations that satisfy the specification's contracts, invariants, and acceptance tests survive validation \cite{piskala2026spec,marri2026constitutional}.

\subsection{Historical Lineages}
\label{sec:lineage}

SDD is not a de novo invention of the LLM era; it adapts several mature research traditions, and recognizing this lineage matters for both intellectual credit and methodological transfer.

\textbf{Program synthesis.} The idea of generating software from specifications has deep roots in deductive synthesis and formal methods. Program \emph{sketching} demonstrated that a programmer-provided partial specification, combined with constraint solving, could automatically complete low-level implementation details, yielding correct-by-construction programs \cite{solarlezama2008program}. Surveying the field, Gulwani et al.\ established that specification-driven techniques can guarantee functional correctness but historically struggled with scalability and expressiveness \cite{gulwani2017program}---precisely the two dimensions on which LLMs excel. SDD can therefore be read as a synthesis of the two traditions: LLM fluency amalgamated with specification rigor.

\textbf{Model-driven engineering.} A second lineage runs through model-driven engineering (MDE), in which models rather than code serve as primary artifacts and transformations derive implementations from them. Among the grand challenges of MDE is the need for AI-augmented model transformations that preserve semantic fidelity from specification to executable system \cite{bucchiarone2020grand}. SDD realizes a version of this program: the specification is an operational artifact from which agents synthesize code, derive tests, and generate deployment descriptors, with the persistent machine-readable specification anchoring design--implementation consistency across iterations.

\textbf{Design by contract.} Meyer introduced the formal specification of software behavior through preconditions, postconditions, and invariants attached to interfaces, arguing that contracts make correctness a checkable property rather than an aspiration \cite{meyer1992applying}. Contemporary SDD extends this tradition directly: contract-like constraints---functional, security, and regulatory---are applied to AI code generation, and CSDD explicitly grounds its constitution in Meyer's contracts \cite{marri2026constitutional}.

\textbf{Requirements economics.} Decades of requirements-engineering research established that early investment in requirements and specification decreases downstream defect and maintenance costs, a classical result whose canonical economic formulation traces to Boehm \cite{boehm1981software}. AI-assisted development strengthens rather than weakens this argument, because LLMs perform substantially better within well-defined contextual boundaries than when inferring unstated assumptions from conversational prompts \cite{zan2023large,piskala2026spec}. The observation also echoes Brooks's distinction between essence and accident \cite{brooks1987no}: generation automates the accidental labor of syntax, leaving the essential complexity of specification untouched---and therefore newly load-bearing.

\subsection{Positioning}
\label{sec:positioning}

Relative to this literature, the present article occupies a distinct position. The empirical studies cited above each examine one facet---productivity \cite{peng2023impact,borg2026echoes}, practitioner behavior \cite{fawzy2026vibe}, security \cite{pearce2022asleep,perry2023users,fu2025security}, test adequacy \cite{liu2023your}---while the methodological treatments \cite{piskala2026spec,marri2026constitutional} articulate SDD from the practitioner's perspective without situating it against the full quality-attribute demands of the enterprise. No prior work, to the best of our knowledge, provides a formal, tool-independent reference model of specification governance together with a systematic, standard-anchored assessment of the empirical evidence for and against it. That is the contribution attempted here.

% ================= METHOD =================
\section{Research Method}
\label{sec:method}

The study follows a design-science structure \cite{hevner2004design,wieringa2014design}: a designed artifact---the Specification Governance Reference Model---is constructed from a body of prior knowledge and evaluated against explicit criteria. Because the evidence base for AI-native development practices is young and spans both peer-reviewed venues and preprints, the knowledge-base construction adapts the guidance of systematic and multivocal review methodology \cite{kitchenham2007guidelines,garousi2019guidelines} with a strict verification discipline. The method comprises four stages.

\textbf{Stage 1: Corpus assembly and verification.} Candidate sources were assembled from (i)~systematic literature reviews of LLM-based software engineering \cite{hou2024large,fan2023large}, (ii)~the citation graphs of the foundational SDD and vibe-coding treatments \cite{piskala2026spec,meske2025vibe,fawzy2026vibe}, and (iii)~targeted searches for empirical studies of AI-assisted development quality, security, and maintainability. Every candidate reference was verified against a primary source---an arXiv abstract page, publisher page, DBLP record, or official proceedings---before inclusion. Claims for which no verifiable source could be located were excluded rather than attributed to unverifiable citations; this discipline is itself a response to a documented failure mode of LLM-assisted scholarship. The resulting corpus comprises 44 verified sources spanning controlled experiments, user studies, grey-literature reviews, benchmark studies, enterprise case studies, methodological treatments, standards, and foundational works (Table~\ref{tab:corpus}).

\begin{table}[htbp]
\centering
\small
\caption{Composition of the verified evidence corpus.}
\label{tab:corpus}
\begin{tabular}{L{5.8cm}cL{6cm}}
\toprule
\rowcolor{rowtint}
\textbf{Category} & \textbf{n} & \textbf{Representative sources} \\
\midrule
Controlled experiments \& user studies & 5 & \cite{peng2023impact,borg2026echoes,perry2023users,sandoval2023lost,liu2023your} \\
Empirical/benchmark studies at scale & 5 & \cite{pearce2022asleep,fu2025security,cotroneo2025human,lukasczyk2023empirical,lemieux2023codamosa} \\
Grey-literature / practitioner evidence & 2 & \cite{fawzy2026vibe,karpathy2025vibe} \\
Enterprise case studies & 2 & \cite{vilasboas2026one,marri2026constitutional} \\
Systematic reviews \& surveys & 5 & \cite{hou2024large,fan2023large,zan2023large,gulwani2017program,alenezi2026rise} \\
Methodological \& vision treatments & 8 & \cite{piskala2026spec,hassan2024towards,alenezi2026rethinking,meske2025vibe,kessel2024nversion,kessel2025morescient,ozkaya2023application,bucchiarone2020grand} \\
Foundational neural code generation & 3 & \cite{chen2021evaluating,li2022competition,karampatsis2020big} \\
Standards \& foundational works & 10 & \cite{iso25010,meyer1992applying,boehm1981software,brooks1987no,parnas1972criteria,sculley2015hidden,solarlezama2008program,clelandhuang2014software,belani2019requirements,bhargavan2016formal} \\
Methodology guidance & 4 & \cite{hevner2004design,wieringa2014design,kitchenham2007guidelines,garousi2019guidelines} \\
\bottomrule
\end{tabular}
\end{table}

\textbf{Stage 2: Failure-mode analysis (RQ1).} The empirical sources were coded for reported quality outcomes of AI-assisted development and grouped into recurring failure modes by the quality attribute affected and the causal mechanism implicated. A failure mode was admitted only if supported by at least two independent verified sources, and counter-evidence was explicitly recorded rather than discarded (Section~\ref{sec:failures}).

\textbf{Stage 3: Reference-model construction (RQ2).} The SGRM was constructed by (i)~formalizing the specification and validation concepts recurring across the SDD literature \cite{piskala2026spec,marri2026constitutional,vilasboas2026one} into definitions with explicit semantics; (ii)~mapping each RQ1 failure mode to the mechanism intended to interdict it; and (iii)~expressing the model's empirical commitments as six falsifiable design propositions. Formalization here serves precision and falsifiability, not mathematical depth: the definitions fix what is meant by specification governance so that the evaluation cannot equivocate.

\textbf{Stage 4: Analytical evaluation (RQ3).} The model is evaluated by \emph{informed argument} \cite{hevner2004design} against the ISO/IEC~25010 product quality model \cite{iso25010}: each quality characteristic is traced to the documented risk it faces under ungoverned generation, the SGRM mechanism addressing that risk, and the published empirical evidence bearing on the mechanism---including null results and unreplicated findings, which are weighted accordingly. No new experimental data are collected; the evaluation's strength derives from the breadth and verification discipline of the corpus, and its limits are examined in Section~\ref{sec:threats}.

% ================= FAILURE MODES =================
\section{Failure Modes of Ungoverned Generation (RQ1)}
\label{sec:failures}

This section consolidates the verified empirical record into four recurring failure modes of ungoverned, conversational AI code generation. ``Ungoverned'' is used precisely: the failure modes attach to generation practices in which acceptance is decided by observational sampling of behavior rather than by verification against an explicit contract---the defining property of vibe coding \cite{fawzy2026vibe,meske2025vibe}---not to AI assistance as such. Table~\ref{tab:failures} summarizes the failure modes, their causal mechanisms, and the supporting evidence; the counter-evidence is treated in Section~\ref{sec:counterevidence}.

\begin{table}[htbp]
\centering
\small
\caption{Failure modes of ungoverned AI code generation identified from the verified evidence corpus.}
\label{tab:failures}
\begin{tabular}{L{4.6cm}L{7.2cm}L{2cm}}
\toprule
\rowcolor{rowtint}
\textbf{Failure mode} & \textbf{Causal mechanism} & \textbf{Key evidence} \\
\midrule
F1: Productivity--reliability paradox & Speed and flow incentivize acceptance on observed behavior; QA practice thin or absent; correctness gap between ``appears to work'' and ``is correct'' & \cite{fawzy2026vibe,liu2023your,cotroneo2025human} \\
\addlinespace
F2: Architectural erosion & Generator optimizes the localized prompt context; cross-cutting system properties are invisible from within a single generation; local errors compound directionally & \cite{fawzy2026vibe,cotroneo2025human,alenezi2026rise,piskala2026spec} \\
\addlinespace
F3: Security exposure & Vulnerability-prone generation compounded by overconfident human acceptance; weaknesses reach production repositories & \cite{pearce2022asleep,perry2023users,sandoval2023lost,fu2025security} \\
\addlinespace
F4: Technical-debt accumulation & Reasoning behind generated code never externalized into reviewable artifacts; entanglement and drift between intent and implementation & \cite{sculley2015hidden,fawzy2026vibe,cotroneo2025human} \\
\bottomrule
\end{tabular}
\end{table}

\subsection{F1: The Productivity--Reliability Paradox}
\label{sec:paradox}

The productivity benefits of AI-assisted development are among the best-replicated findings in the recent empirical literature. In a controlled experiment with GitHub Copilot, developers with AI assistance completed a standardized implementation task approximately 55.8\% faster than a control group \cite{peng2023impact}. A preregistered two-phase controlled experiment with 151 participants (95\% professional developers) observed a 30.7\% median reduction in completion time for AI-assisted feature development, with habitual AI users showing an estimated 55.9\% speedup \cite{borg2026echoes}. Large-scale surveys document substantial implementation-phase gains across the lifecycle \cite{hou2024large,fan2023large}.

Against this promise stands what Fawzy et al.\ empirically identify as a speed--quality trade-off paradox \cite{fawzy2026vibe}. Their systematic grey-literature review of 101 practitioner sources, from which 518 firsthand behavioral accounts were extracted, found that vibe coders are motivated primarily by speed and accessibility and frequently experience rapid ``instant success'' and flow---yet most perceive the resulting code quality as unreliable, and quality-assurance practice is thin or absent. A common experience is unstructured prompt iteration: cycles of prompt adjustment and code refinement, sometimes requiring dozens or hundreds of iterations, substituting trial-and-error for systematic design \cite{fawzy2026vibe}. Related work sharpens the reliability side of the paradox. The EvalPlus framework demonstrated that many LLM-generated solutions passing the simple test suites of standard benchmarks fail under rigorously augmented, specification-derived test suites, substantially reducing measured pass rates and exposing a \emph{correctness gap} between apparent and actual correctness \cite{liu2023your}. A large-scale comparison of human-written and AI-generated code documents systematic differences across defects, vulnerabilities, and complexity that unreviewed acceptance would import directly into a codebase \cite{cotroneo2025human}. The verified record thus supports a precise proposition: unreviewed conversational generation is empirically associated with defects and vulnerabilities that simple local execution does not reveal, and its immediate productivity gains are offset by downstream quality liabilities \cite{fawzy2026vibe,liu2023your,pearce2022asleep,cotroneo2025human}.

\subsection{F2: Cognitive Limits of Localized Generation and Architectural Erosion}
\label{sec:erosion}

The fundamental technical vulnerability of conversational generation lies in the mismatch between a language model's localized context and the global requirements of a large system. LLMs excel at optimizing isolated snippets and self-contained modules; they are far weaker at evaluating multi-layered architectural trade-offs---horizontal scaling, cache consistency, database replication, cross-service transactional integrity---that span a distributed system and are invisible from within a single prompt's context \cite{alenezi2026rise,piskala2026spec}. Representational capacity is not architectural judgment: models can learn rich representations over large codebases \cite{karampatsis2020big}, yet because generation optimizes the localized context it is given, conversational workflows without systemic framing tend to introduce duplicated logic, missing input validation, and subtle logical defects that escape local execution \cite{fawzy2026vibe,cotroneo2025human}.

Over time, localized errors compound. Whereas individual generation errors behave like random noise, the acceptance of AI-generated code without checking its systemic impact introduces architectural fragmentation that conflicts with broader enterprise patterns and hinders future refactoring \cite{fawzy2026vibe}---a directional decay consistent with the long-established phenomenon of architectural erosion. This failure mode has a classical explanation: modularity criteria and information hiding are \emph{system-level} design decisions \cite{parnas1972criteria}, and quality attributes are emergent, cross-cutting properties; no sequence of locally optimal generations converges on them by accident. The practical consequence is that casual natural-language prompting works well for standalone utilities and initial experiments but is insufficient for complex systems, which require an explicit, durable representation of the architecture against which every generated increment can be checked. Providing that representation is precisely the function of specification governance (Section~\ref{sec:sgrm}).

\subsection{F3: Security Exposure}
\label{sec:security}

Security is the domain in which the evidence against unstructured generation is most direct and most quantified. In the landmark ``Asleep at the Keyboard?'' study, GitHub Copilot was evaluated across scenarios relevant to MITRE's high-risk weakness categories; approximately 40\% of the generated programs contained exploitable vulnerabilities \cite{pearce2022asleep}. The risk propagates through the human in the loop: participants with access to an AI assistant wrote significantly less secure code than controls while being \emph{more} likely to believe their code was secure \cite{perry2023users}, and a security-focused user study of LLM code assistants documented the conditions under which assistant use affects vulnerability introduction in low-level code \cite{sandoval2023lost}. At repository scale, an empirical analysis of security weaknesses in Copilot-generated code found in real GitHub projects confirms that benchmark-measured vulnerability rates materialize in production artifacts \cite{fu2025security}, and the large-scale human-versus-AI comparison extends the picture across defect and vulnerability classes \cite{cotroneo2025human}. The combination is what makes this failure mode sharp: vulnerability-prone generation \emph{and} overconfident acceptance jointly imply that, absent structural controls, AI-generated code accepted without systematic review imports security debt.

\subsection{F4: Technical-Debt Accumulation}
\label{sec:debt}

Maintainability is the dimension along which the difference between governed and ungoverned generation compounds over time. The foundational warning predates LLMs: machine-learning systems accumulate \emph{hidden technical debt}---entanglement, undeclared dependencies, configuration sprawl, glue code---whose carrying cost dwarfs initial development savings \cite{sculley2015hidden}. The lesson transfers directly: quick-prompted functions can entangle logic in ways that make future change costly, precisely because the reasoning that produced them was never externalized into a reviewable artifact. The vibe-coding literature supplies contemporary evidence: practitioners themselves perceive vibe-coded output as unreliable and hard to maintain, with quality assurance deferred or skipped, and debt accumulation is a core theme of the extracted behavioral accounts \cite{fawzy2026vibe}; measured complexity and defect differences between human-written and AI-generated code corroborate the mechanism at scale \cite{cotroneo2025human}.

\subsection{Counter-Evidence and the Moderating Variable}
\label{sec:counterevidence}

Intellectual honesty requires recording a significant counterpoint. Borg et al., in the controlled experiment cited above, did \emph{not} detect systematic code-level maintainability disadvantages when new developers later evolved code co-developed with AI assistants, within the scope of their tasks and measures \cite{borg2026echoes}. This preregistered result cautions against overclaiming that AI assistance \emph{per se} degrades maintainability. The reconciliation of the two bodies of evidence is itself informative and motivates the reference model: the experimental conditions in \cite{borg2026echoes} involved conventional professional review discipline, whereas the debt evidence attaches to \emph{unstructured, unreviewed} generation \cite{fawzy2026vibe}. The moderating variable, on the present reading, is not the presence of AI but the presence of \emph{governance}: AI assistance embedded in disciplined engineering practice appears maintainability-neutral or better, while AI generation substituting for disciplined practice accumulates debt. This reading is a hypothesis consistent with both datasets rather than a directly tested finding---a status revisited in Sections~\ref{sec:discussion} and~\ref{sec:threats}. Open risks flagged by the experimenters---code bloat from excessive generation and cognitive debt from mental offloading---remain open empirical questions \cite{borg2026echoes}.

\textbf{Answer to RQ1.} The verified evidence identifies four recurring failure modes of ungoverned AI code generation---the productivity--reliability paradox (F1), architectural erosion (F2), security exposure (F3), and technical-debt accumulation (F4)---sharing a common causal structure: a stochastic generator optimizing localized context, coupled to an acceptance criterion (observed behavior) that is strictly weaker than the quality attributes the enterprise requires.

% ================= REFERENCE MODEL =================
\section{The Specification Governance Reference Model (RQ2)}
\label{sec:sgrm}

The common causal structure of the RQ1 failure modes suggests a common remedy: replace acceptance-by-observation with acceptance-by-verification against an explicit, durable contract. This section consolidates that remedy into the \emph{Specification Governance Reference Model} (SGRM), a tool-independent formalization of specification-driven development synthesized from the methodological literature \cite{piskala2026spec,marri2026constitutional,vilasboas2026one} and the historical lineages of Section~\ref{sec:lineage}. The model comprises formal definitions (Section~\ref{sec:formal}), a rigor-tier formalization (Section~\ref{sec:tiers}), a layered architecture (Section~\ref{sec:layers}), a closed-loop regeneration algorithm (Section~\ref{sec:loop}), and six falsifiable design propositions (Section~\ref{sec:propositions}).

\subsection{Formal Model}
\label{sec:formal}

\begin{definition}[Specification]
\label{def:spec}
A specification is a machine-readable tuple $\spec = (F, Q, K, \Sigma)$, where $F$ is a set of \emph{functional obligations} (interface signatures with preconditions, postconditions, and invariants in the sense of design by contract \cite{meyer1992applying}, acceptance scenarios, and behavioral examples); $Q$ is a set of \emph{quality constraints} (measurable thresholds over the quality characteristics of ISO/IEC~25010 \cite{iso25010}, e.g., latency budgets, availability targets, complexity bounds); $K$ is a set of \emph{constitutional constraints} (non-negotiable security, privacy, and regulatory rules, e.g., prohibitions derived from CWE weakness classes \cite{marri2026constitutional}); and $\Sigma$ is an \emph{architectural structure} (module boundaries, bounded contexts, interface contracts, and dependency rules \cite{parnas1972criteria}).
\end{definition}

\begin{definition}[Generator]
\label{def:generator}
A generator is a stochastic function $\gen : (\spec, C) \mapsto \impl$ that, given a specification $\spec$ and a context $C$ (codebase excerpts, diagnostics, conversation history), samples an implementation artifact $\impl$ from a probability distribution $p_\gen(\impl \mid \spec, C)$. Nondeterminism is intrinsic: repeated invocation with identical inputs yields materially different artifacts \cite{meske2025vibe,zan2023large}.
\end{definition}

\begin{definition}[Validator]
\label{def:validator}
A validator is a deterministic, decidable procedure $\val(\spec, \impl) \in \{\top, \bot\}$ decomposed as
$\val = \val_{\mathrm{stat}} \wedge \val_{\mathrm{test}} \wedge \val_{\mathrm{contract}} \wedge \val_{\mathrm{arch}}$,
where $\val_{\mathrm{stat}}$ applies static analysis and constitutional checks against $K$; $\val_{\mathrm{test}}$ executes test suites derived from $F$ (including specification-derived augmentation in the sense of EvalPlus \cite{liu2023your}); $\val_{\mathrm{contract}}$ checks interface contracts and, where $F$ admits formal semantics, invokes proof obligations \cite{bhargavan2016formal}; and $\val_{\mathrm{arch}}$ checks conformance of dependencies and module boundaries against $\Sigma$. The \emph{acceptance set} of a specification is $\accept(\spec) = \{\impl \mid \val(\spec,\impl) = \top\}$.
\end{definition}

\begin{definition}[Specification governance]
\label{def:governance}
A development process is \emph{specification-governed} iff (i)~every artifact integrated into the system belongs to $\accept(\spec)$ for the current specification $\spec$, and (ii)~every intended change to system behavior is initiated by a change to $\spec$, never by direct mutation of $\impl$ alone.
\end{definition}

This vocabulary makes the contrast with vibe coding precise. In vibe coding, the specification is \emph{latent}: it exists only as an implicit intent distributed across working memory and conversation history \cite{meske2025vibe,fawzy2026vibe}. Since no explicit $\spec$ exists, $\accept(\spec)$ is undefined, and acceptance degenerates to \emph{observational sampling}---executing the artifact on a handful of inputs and judging the ``vibe'' of the result. Observational sampling is a strictly weaker acceptance criterion than membership in $\accept(\spec)$: it examines finitely many behaviors of a single sampled artifact, whereas the validator checks obligations quantified over $F$, $Q$, $K$, and $\Sigma$. Every RQ1 failure mode exploits this gap: F1 because sampled behavior conceals the correctness gap \cite{liu2023your}; F2 because $\Sigma$ is never represented, so erosion is invisible; F3 because $K$ is never checked and human confidence substitutes for verification \cite{perry2023users}; F4 because intent is never externalized, so drift between intent and implementation cannot be detected, let alone repaired.

\subsection{Rigor Tiers}
\label{sec:tiers}

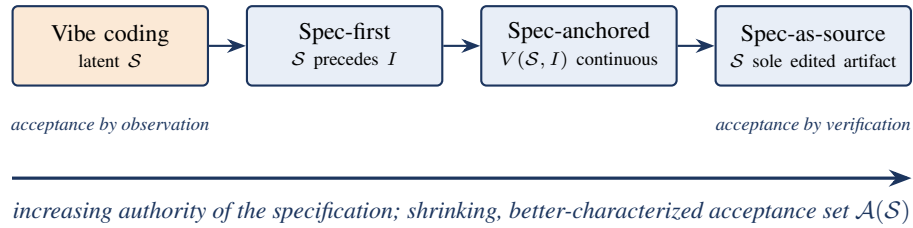
\begin{figure}[htbp]
\centering
\begin{tikzpicture}[
  every node/.style={font=\small},
  stage/.style={draw=tableheader, thick, rounded corners=2pt, fill=boxtint,
                minimum height=1.05cm, text width=2.35cm, align=center},
  gstage/.style={draw=tableheader, thick, rounded corners=2pt, fill=gatetint,
                minimum height=1.05cm, text width=2.35cm, align=center}
]
  \node[gstage] (vibe) at (0,0) {Vibe coding\\ \scriptsize latent $\spec$};
  \node[stage] (first) at (3.1,0) {Spec-first\\ \scriptsize $\spec$ precedes $\impl$};
  \node[stage] (anch) at (6.2,0) {Spec-anchored\\ \scriptsize $\val(\spec,\impl)$ continuous};
  \node[stage] (source) at (9.3,0) {Spec-as-source\\ \scriptsize $\spec$ sole edited artifact};
  \draw[-{Stealth[length=2.5mm]}, thick, tableheader] (vibe) -- (first);
  \draw[-{Stealth[length=2.5mm]}, thick, tableheader] (first) -- (anch);
  \draw[-{Stealth[length=2.5mm]}, thick, tableheader] (anch) -- (source);
  \node at (0,-1.05) [font=\scriptsize\itshape, text=tableheader]{acceptance by observation};
  \node at (9.3,-1.05) [font=\scriptsize\itshape, text=tableheader]{acceptance by verification};
  \draw[-{Stealth[length=3mm]}, very thick, tableheader]
      (-1.3,-1.75) -- (10.6,-1.75)
      node[midway, below=1.5mm, font=\small\itshape]
      {increasing authority of the specification; shrinking, better-characterized acceptance set $\accept(\spec)$};
\end{tikzpicture}
\caption{The rigor continuum from vibe coding to spec-as-source development. Moving right, the specification gains authority over the implementation and the acceptance criterion strengthens from observational sampling to verified membership in $\accept(\spec)$.}
\label{fig:continuum}
\end{figure}

Piskala's maturity model \cite{piskala2026spec} distinguishes three tiers of specification rigor, which the SGRM formalizes as progressively stronger process invariants over the pair $(\spec, \impl)$ (Table~\ref{tab:tiers}, Figure~\ref{fig:continuum}).

\begin{table}[htbp]
\centering
\small
\caption{The three rigor tiers of specification-driven development, formalized as process invariants. Tiers follow Piskala \cite{piskala2026spec}; the invariant formulation is contributed here.}
\label{tab:tiers}
\begin{tabular}{L{2cm}L{3.2cm}L{4.8cm}L{3.8cm}}
\toprule
\rowcolor{rowtint}
\textbf{Tier} & \textbf{Human-edited artifact} & \textbf{Process invariant} & \textbf{Maintenance semantics} \\
\midrule
Spec-first & $\spec$ then $\impl$ & $\spec$ authored before $\impl$; $\val(\spec,\impl)=\top$ at integration time only & $\impl$ evolves independently; the $\spec$--$\impl$ bond may decay \\
\addlinespace
Spec-anchored & $\spec$ and $\impl$ & $\val(\spec,\impl)=\top$ maintained \emph{continuously}; any edit to either artifact re-triggers validation & Divergence detected immediately; spec and code co-evolve \\
\addlinespace
Spec-as-source & $\spec$ only & Human edit operations restricted to $\spec$; $\impl$ regenerated s.t.\ $\impl \in \accept(\spec)$ & Maintenance $=$ specification change $+$ regeneration \\
\bottomrule
\end{tabular}
\end{table}

The tiers are not a ranking of virtue but a costed design space: each step rightward buys stronger guarantees at higher specification-authoring and tooling cost, and Piskala's decision framework makes tier selection an explicit engineering trade-off \cite{piskala2026spec}. The evaluation in Section~\ref{sec:evaluation} shows that the enterprise quality attributes load principally on the spec-anchored invariant---continuous validation---while spec-as-source additionally transforms the economics of maintenance (Section~\ref{sec:evalmaint}).

\subsection{Layered Architecture}
\label{sec:layers}

\begin{figure}[htbp]
\centering
\begin{tikzpicture}[
  every node/.style={font=\small},
  layer/.style={draw=tableheader, thick, rounded corners=3pt, fill=boxtint,
                minimum height=1.5cm, text width=9.2cm, align=left, inner sep=7pt},
  comp/.style={font=\scriptsize},
  spine/.style={draw=tableheader, thick, rounded corners=3pt, fill=gatetint,
                minimum width=2.1cm, align=center, inner sep=5pt}
]
  \node[layer] (l1) at (0,0)
    {\textbf{L1 Specification layer}\\[1pt]
     \begin{minipage}{8.8cm}\scriptsize
     $\spec=(F,Q,K,\Sigma)$: contracts, scenarios; quality thresholds (ISO/IEC 25010); constitutional security \& regulatory rules; architectural structure. Version-controlled; the sole source of truth.
     \end{minipage}};
  \node[layer, below=0.42cm of l1] (l2)
    {\textbf{L2 Generation layer}\\[1pt]
     \begin{minipage}{8.8cm}\scriptsize
     Stochastic generators $\gen(\spec,C)$: LLMs and specialized agents (implementation, test, review, compliance) planning and synthesizing candidate artifacts within the guardrails of $\spec$.
     \end{minipage}};
  \node[layer, below=0.42cm of l2] (l3)
    {\textbf{L3 Verification layer}\\[1pt]
     \begin{minipage}{8.8cm}\scriptsize
     Deterministic validator $\val=\val_{\mathrm{stat}}\wedge\val_{\mathrm{test}}\wedge\val_{\mathrm{contract}}\wedge\val_{\mathrm{arch}}$: static \& constitutional analysis, specification-derived tests, contract/proof obligations, architectural conformance; N-version differential assessment.
     \end{minipage}};
  \node[layer, below=0.42cm of l3] (l4)
    {\textbf{L4 Governance layer}\\[1pt]
     \begin{minipage}{8.8cm}\scriptsize
     Human oversight and accountability: specification review boards, integration gates, audit trails, compliance reporting, escalation on validation exhaustion.
     \end{minipage}};
  \node[spine, fit={($(l1.north east)+(0.28,0)$) ($(l4.south east)+(2.35,0)$)},
        label={[font=\scriptsize\bfseries, rotate=90, anchor=south]center:Traceability spine \; $T \subseteq \mathrm{elem}(\spec)\times \mathrm{artifacts}$}] (spine) {};
  \draw[-{Stealth[length=2.5mm]}, thick, tableheader] (l1) -- node[right, comp]{generate} (l2);
  \draw[-{Stealth[length=2.5mm]}, thick, tableheader] (l2) -- node[right, comp]{candidates} (l3);
  \draw[-{Stealth[length=2.5mm]}, thick, tableheader] (l3) -- node[right, comp]{verified artifacts / escalations} (l4);
  \draw[-{Stealth[length=2.5mm]}, thick, tableheader, dashed]
       ($(l3.west)+(0,0)$) .. controls ($(l3.west)+(-1.1,0)$) and ($(l2.west)+(-1.1,0)$) ..
       ($(l2.west)+(0,0)$);
  \node[comp, align=center, text=tableheader, rotate=90, fill=white, inner sep=1pt]
       at ($(l2.west)!0.5!(l3.west)+(-1.12,0)$) {diagnostics as\\ corrective context};
  \draw[-{Stealth[length=2.5mm]}, thick, tableheader, dashed]
       ($(l4.west)+(0,0)$) .. controls ($(l4.west)+(-2.6,0)$) and ($(l1.west)+(-2.6,0)$) ..
       ($(l1.west)+(0,0)$);
  \node[comp, align=center, text=tableheader, rotate=90, fill=white, inner sep=1pt]
       at ($(l1.west)!0.5!(l4.west)+(-2.55,0)$) {specification\\ evolution};
\end{tikzpicture}
\caption{The Specification Governance Reference Model (SGRM). Solid arrows denote the generation-and-verification flow; dashed arrows denote feedback (validator diagnostics as corrective context; governed evolution of the specification). The traceability spine links every specification element to the artifacts that realize it.}
\label{fig:sgrm}
\end{figure}
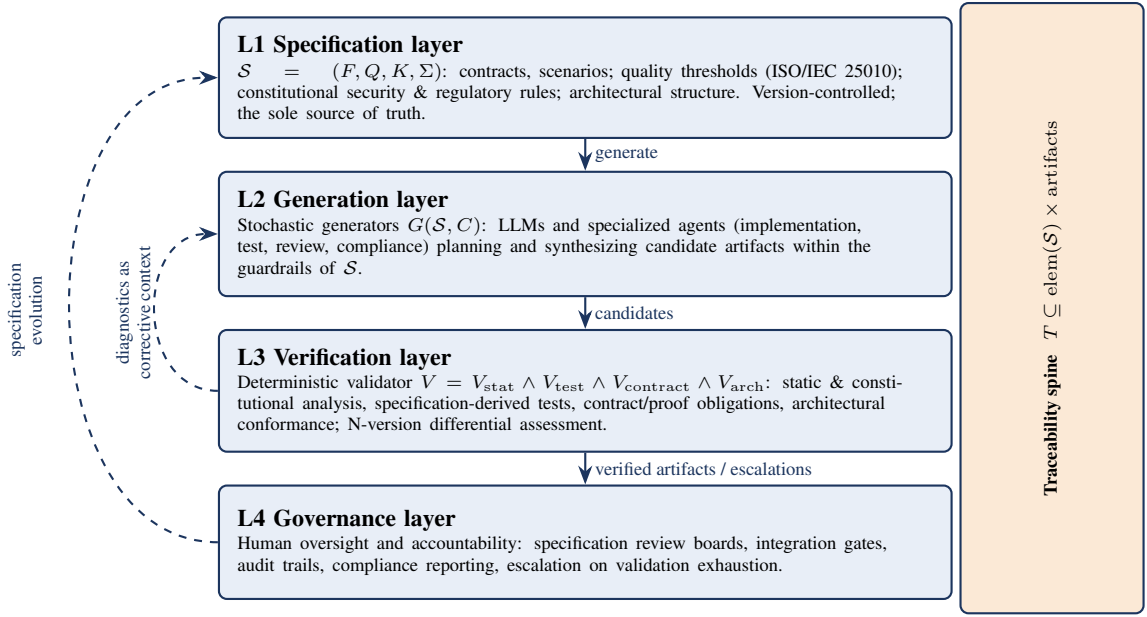

Figure~\ref{fig:sgrm} organizes the model into four horizontal layers crossed by a vertical traceability spine.

\textbf{L1: Specification layer.} The specification $\spec$ of Definition~\ref{def:spec} is the version-controlled source of truth. Its four components are jointly authored by humans (with AI assistance) but are the \emph{only} artifacts whose semantics bind the system. Externalizing intent here is what converts ``the model misunderstood me'' from an unfalsifiable complaint into a checkable divergence between artifact and contract \cite{piskala2026spec,clelandhuang2014software}.

\textbf{L2: Generation layer.} One or more stochastic generators---general-purpose LLMs or specialized agents for implementation, test synthesis, review, and compliance \cite{vilasboas2026one}---produce candidate artifacts from $\spec$ and context. The layer is deliberately agnostic about models and tools; the reference model constrains only the interface: generators consume specifications and diagnostics, and emit candidates.

\textbf{L3: Verification layer.} The validator of Definition~\ref{def:validator} is the deterministic boundary around the stochastic core. Its components instantiate verified techniques: specification-derived test augmentation exposes the correctness gap \cite{liu2023your}; automated unit-test generation supplies coverage machinery \cite{lukasczyk2023empirical}, with LLM assistance escaping coverage plateaus when guided by clear context \cite{lemieux2023codamosa}; constitutional checks enforce $K$ mechanically \cite{marri2026constitutional}; formal contract obligations, where applicable, exclude entire vulnerability classes \cite{bhargavan2016formal}; and N-version differential assessment exploits generation itself, evaluating quality through comparative analysis across independently sampled versions so that reliability rests on version diversity rather than trust in any single artifact \cite{kessel2024nversion}.

\textbf{L4: Governance layer.} Humans retain accountability: specification review, integration gates, audit trails, and escalation when the regeneration budget is exhausted. Traceability is restored \emph{by construction}: because every artifact enters the system through validation against $\spec$, the relation $T$ linking specification elements to realizing artifacts is a byproduct of the process rather than an after-the-fact reconstruction---inverting the classical economics of traceability, in which links are recovered manually at high cost \cite{clelandhuang2014software}. A persistent specification artifact is likewise what certification regimes in safety-critical domains require \cite{belani2019requirements}.

\subsection{The Closed Loop}
\label{sec:loop}

Algorithm~\ref{alg:loop} states the operational core of the model: specification-governed generation is a rejection-sampling loop in which the deterministic validator filters the stochastic generator's proposals, and validation diagnostics feed back as corrective context.

\begin{algorithm}[htbp]
\caption{Closed-loop specification-governed generation}
\label{alg:loop}
\begin{algorithmic}[1]
\REQUIRE specification $\spec = (F, Q, K, \Sigma)$; generator $\gen$; validator $\val$; budget $B$
\ENSURE verified artifact $\impl \in \accept(\spec)$, or escalation to L4
\STATE $C \leftarrow$ initial context (codebase excerpts, plan derived from $\spec$)
\FOR{$i = 1$ \TO $B$}
  \STATE sample candidate $\impl_i \sim p_\gen(\cdot \mid \spec, C)$
  \STATE $r \leftarrow \val(\spec, \impl_i)$ \COMMENT{static $\wedge$ test $\wedge$ contract $\wedge$ architectural conformance}
  \IF{$r = \top$}
    \STATE record traceability links $T \leftarrow T \cup \{(e, \impl_i) : e \in \mathrm{elem}(\spec) \text{ realized by } \impl_i\}$
    \RETURN $\impl_i$ \COMMENT{integrate}
  \ELSE
    \STATE $C \leftarrow C \cup \mathrm{diagnostics}(r)$ \COMMENT{failures become corrective context}
  \ENDIF
\ENDFOR
\STATE \textbf{escalate} to governance layer \COMMENT{human revises $\spec$, decomposes the task, or implements manually}
\end{algorithmic}
\end{algorithm}

Three properties of the loop deserve emphasis. First, it is \emph{self-correcting by design}: the same specification that drives generation also generates the oracle against which candidates are judged, closing the loop that vibe coding leaves open \cite{piskala2026spec,marri2026constitutional,vilasboas2026one}. Second, it is \emph{monotone in the validator}: strengthening any component of $\val$ shrinks $\accept(\spec)$ and thus tightens the guarantee attached to every accepted artifact, without touching the generator---quality improvements compose through the deterministic boundary rather than through model retraining. Third, it \emph{degrades safely}: budget exhaustion escalates to human governance rather than silently lowering the bar. The loop presupposes semantic limitations of current generators rather than wishing them away: because code models are trained overwhelmingly on the syntactic facet of software, their trustworthiness is lowest exactly where semantics matter \cite{kessel2025morescient}; until semantically grounded (``Morescient'') models mature, the gap between syntactic plausibility and semantic adequacy must be closed by process---by $\val$---rather than assumed away.

\subsection{Design Propositions}
\label{sec:propositions}

The SGRM's empirical commitments are summarized as six falsifiable design propositions, evaluated against the evidence in Section~\ref{sec:evaluation}.

\begin{itemize}
\item \textbf{DP1 (Determinism boundary).} Bounding acceptance by a deterministic validator converts stochastic generation into a contract-based workflow whose output variance, over the properties expressed in $\spec$, is limited to $\accept(\spec)$.
\item \textbf{DP2 (Reproducibility).} An explicit $\spec$ makes AI-assisted development reproducible and cumulative across sessions, team members, and model versions, where conversational prompting is none of these.
\item \textbf{DP3 (Verification by construction).} Deriving tests, contracts, and constitutional checks from $\spec$ yields systematically stronger defect and vulnerability detection than observational acceptance.
\item \textbf{DP4 (Architectural integrity).} Representing $\Sigma$ explicitly and checking $\val_{\mathrm{arch}}$ continuously interdicts the erosion mechanism of F2.
\item \textbf{DP5 (Maintenance economics).} At the spec-anchored tier and above, maintenance cost shifts from code archaeology to specification evolution; at spec-as-source, regeneration bounded by $\val$ acts as continuous refactoring at low marginal human cost.
\item \textbf{DP6 (Traceability and compliance).} Specification governance yields audit-ready traceability as a process byproduct, reducing compliance demonstration to two checks: that $\spec$ captures the applicable obligations, and that the pipeline enforces $\val$.
\end{itemize}

\textbf{Answer to RQ2.} Specification governance addresses the RQ1 failure modes through a single structural intervention with four faces: externalizing intent as a machine-readable contract (interdicting F4), including architectural structure in that contract (interdicting F2), deriving verification from the contract (interdicting F1), and making constitutional constraints non-negotiable inputs to a deterministic validator (interdicting F3). The SGRM consolidates these mechanisms into a tool-independent reference model whose commitments are stated as falsifiable propositions.

% ================= EVALUATION =================
\section{Evidence-Based Evaluation Against ISO/IEC 25010 (RQ3)}
\label{sec:evaluation}

This section evaluates the SGRM analytically against the product quality model of ISO/IEC~25010 \cite{iso25010}, which codifies the characteristics---functional suitability, performance efficiency, compatibility, interaction capability, reliability, security, maintainability, flexibility, and safety---against which enterprise software is assessed. The evaluation proceeds in four steps: a systematic mapping from quality characteristics to risks, mechanisms, and evidence (Section~\ref{sec:mapping}); an examination of the strongest quantified evidence, including the enterprise case studies (Section~\ref{sec:cases}); an explicit weighting of counter-evidence and evidence maturity (Section~\ref{sec:weighting}); and the boundary conditions of the resulting claim (Section~\ref{sec:boundary}). Passing unit tests alone does not qualify software as enterprise-grade; an implementation functionally correct in the narrow sense of satisfying happy-path tests may fail every systemic criterion of the standard. AI-generated code exhibits a characteristic profile in this respect: it frequently succeeds at localized functionality while omitting architectural concerns---dependency management, structured logging, resilience patterns, concurrency control, operational monitoring---unless these are explicitly demanded \cite{fawzy2026vibe,alenezi2026rise}. Under the SGRM, the natural place to demand them is $Q$ and $\Sigma$, where quality constraints sit alongside functional requirements as first-class, machine-checkable obligations.

\subsection{Systematic Mapping to ISO/IEC 25010}
\label{sec:mapping}

Table~\ref{tab:iso} presents the mapping. Each row records a quality characteristic, the documented risk it faces under ungoverned generation (with the failure mode from Section~\ref{sec:failures}), the SGRM mechanism that addresses the risk (with the design proposition it instantiates), and the verified evidence bearing on the mechanism.

\begin{table}[htbp]
\centering
\scriptsize
\caption{Analytical evaluation of the SGRM against the ISO/IEC 25010 product quality characteristics \cite{iso25010}.}
\label{tab:iso}
\begin{tabular}{L{2.1cm}L{4cm}L{4.6cm}L{2cm}}
\toprule
\rowcolor{rowtint}
\textbf{Characteristic} & \textbf{Risk under ungoverned generation} & \textbf{SGRM mechanism} & \textbf{Evidence} \\
\midrule
Functional suitability & Correctness gap: passes casual tests, fails rigorous suites (F1) & Specification-derived test augmentation and contract checks in $\val_{\mathrm{test}}$, $\val_{\mathrm{contract}}$ (DP3) & \cite{liu2023your,lemieux2023codamosa,lukasczyk2023empirical} \\
\addlinespace
Security & $\sim$40\% vulnerable generations; overconfident acceptance; weaknesses in production repos (F3) & Constitutional constraints $K$ enforced in $\val_{\mathrm{stat}}$; security by construction; formal exclusion of vulnerability classes (DP1, DP3) & \cite{pearce2022asleep,perry2023users,fu2025security,marri2026constitutional,bhargavan2016formal} \\
\addlinespace
Maintainability & Debt accumulation; intent never externalized; drift (F4) & $\spec$ as durable intent record; spec-anchored continuous validation; regeneration as continuous refactoring (DP2, DP5) & \cite{sculley2015hidden,fawzy2026vibe,borg2026echoes,piskala2026spec} \\
\addlinespace
Performance efficiency \& scalability & Locally correct artifacts failing under load; cross-cutting properties unrepresented (F2) & Quality thresholds in $Q$; architectural structure $\Sigma$ with conformance checking $\val_{\mathrm{arch}}$ (DP4) & \cite{alenezi2026rise,piskala2026spec,parnas1972criteria,karampatsis2020big} \\
\addlinespace
Reliability & Latent defects invisible to observational sampling (F1) & Layered verification; N-version differential assessment across generated versions (DP1, DP3) & \cite{cotroneo2025human,kessel2024nversion,liu2023your} \\
\addlinespace
Compatibility \& interoperability & Interface drift across conversationally generated components (F2, F4) & API contracts in $F$ validated continuously; single shared referent for humans and agents (DP2, DP4) & \cite{piskala2026spec,vilasboas2026one} \\
\addlinespace
Flexibility (incl.\ adaptability) & Fragmented, tightly coupled implementations resisting change (F2) & Modular $\Sigma$; regeneration on specification change (DP4, DP5) & \cite{parnas1972criteria,piskala2026spec} \\
\addlinespace
Safety \& compliance (context of use) & No navigable chain from regulation to code; decisions buried in chat histories (F4) & Traceability spine $T$ by construction; audit reduces to spec coverage $+$ pipeline enforcement (DP6) & \cite{clelandhuang2014software,belani2019requirements,marri2026constitutional} \\
\bottomrule
\end{tabular}
\end{table}

Three observations follow from the mapping. First, the coverage is structural, not incidental: because $\spec$ has a component for each family of obligations ($F$, $Q$, $K$, $\Sigma$), each characteristic of the standard has a designated place to be \emph{demanded} and a designated validator component to be \emph{checked}. Second, the mechanisms compose: the same closed loop (Algorithm~\ref{alg:loop}) discharges all rows, so adopting the model for one attribute yields the machinery for the others. Third, the evidence column is heterogeneous in maturity---a point developed in Section~\ref{sec:weighting}.

\subsection{Quantified Evidence and Enterprise Case Studies}
\label{sec:cases}

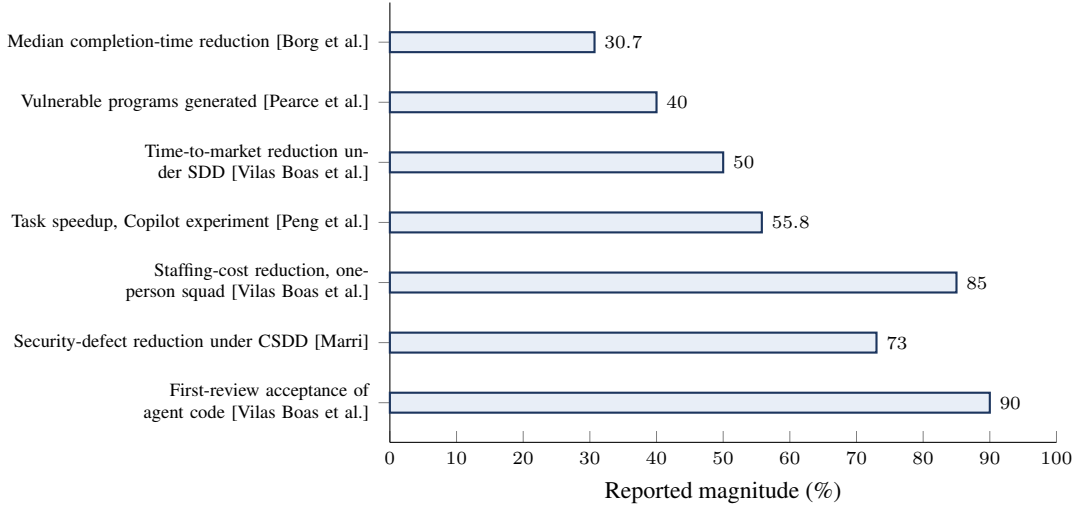
\begin{figure}[htbp]
\centering
\begin{tikzpicture}
\begin{axis}[
  xbar,
  width=10.4cm,
  height=7.4cm,
  xmin=0, xmax=100,
  xlabel={\small Reported magnitude (\%)},
  symbolic y coords={
    {First-review acceptance of agent code [Vilas Boas et al.]},
    {Security-defect reduction under CSDD [Marri]},
    {Staffing-cost reduction, one-person squad [Vilas Boas et al.]},
    {Task speedup, Copilot experiment [Peng et al.]},
    {Time-to-market reduction under SDD [Vilas Boas et al.]},
    {Vulnerable programs generated [Pearce et al.]},
    {Median completion-time reduction [Borg et al.]}},
  ytick=data,
  yticklabel style={font=\scriptsize, align=right, text width=5.1cm},
  xticklabel style={font=\scriptsize},
  bar width=7.5pt,
  axis lines*=left,
  clip=false,
  nodes near coords,
  nodes near coords style={font=\scriptsize, /pgf/number format/fixed},
  enlarge y limits=0.11
]
\addplot[fill=boxtint, draw=tableheader, thick] coordinates {
  (30.7,{Median completion-time reduction [Borg et al.]})
  (40,{Vulnerable programs generated [Pearce et al.]})
  (50,{Time-to-market reduction under SDD [Vilas Boas et al.]})
  (55.8,{Task speedup, Copilot experiment [Peng et al.]})
  (85,{Staffing-cost reduction, one-person squad [Vilas Boas et al.]})
  (73,{Security-defect reduction under CSDD [Marri]})
  (90,{First-review acceptance of agent code [Vilas Boas et al.]})
};
\end{axis}
\end{tikzpicture}
\caption{Selected quantified findings from the verified corpus \cite{peng2023impact,borg2026echoes,pearce2022asleep,marri2026constitutional,vilasboas2026one}. Metrics are heterogeneous across studies (speedups, defect rates, reductions) and are shown together only as an evidence map. The Pearce et al.\ bar is a \emph{risk} finding for ungoverned generation; the Marri and Vilas Boas et al.\ bars are outcomes reported \emph{under} specification governance in single case studies (Section~\ref{sec:weighting}).}
\label{fig:evidence}
\end{figure}

Figure~\ref{fig:evidence} summarizes the strongest quantified findings from the corpus. The figure juxtaposes heterogeneous metrics from heterogeneous designs and is therefore an evidence map, not a meta-analysis; each bar must be interpreted within its study's scope.

\textbf{Security under constitutional governance.} Marri's CSDD case study is the most direct quantified test of DP1 and DP3 in the security dimension: in a banking-microservices application addressing ten critical CWE classes with full traceability from constitutional principles to code locations, constitutional constraints reduced security defects by 73\% relative to unconstrained AI generation while maintaining developer velocity \cite{marri2026constitutional}. The result instantiates the general pattern the formal-methods lineage predicts: when specifications are expressed in a form automated checkers can enforce, entire vulnerability classes become excludable rather than merely detectable \cite{bhargavan2016formal}. Read jointly with the F3 evidence---vulnerability-prone generation \cite{pearce2022asleep,fu2025security} and overconfident acceptance \cite{perry2023users}---the case supports the proposition that proactive security specification outperforms reactive security verification in AI-assisted workflows.

\textbf{Enterprise delivery under specification governance.} Vilas Boas et al.\ report a practitioner-researcher case study at a large Brazilian financial institution---a regulated, brownfield environment---in which a single staff engineer, supported by four specialized AI agents operating under an SDD workflow, delivered a product initiative originally scoped for a four-person cross-functional squad: delivery in half the planned time, a 90\% first-review acceptance rate for AI-generated code, full integration-test pass rates, and an above-85\% reduction in direct staffing cost \cite{vilasboas2026one}. Two interpretive findings matter more than the headline multipliers. First, the binding constraint on AI-augmented delivery was not model capability but the directing engineer's institutional knowledge and the quality of specifications produced upstream---direct enterprise confirmation that under specification governance, the bottleneck of software velocity migrates from generation to specification clarity (DP2). Second, the most consistent gain came from collapsing the outer loop of inter-discipline coordination: with a shared specification as single referent, the communication overhead conventional squads spend aligning requirements, architecture, implementation, testing, and compliance largely disappears \cite{vilasboas2026one}---the organizational-scale reading of DP2 and DP4.

\subsection{Maintainability and the Economics of Regeneration}
\label{sec:evalmaint}

The maintainability row of Table~\ref{tab:iso} deserves elaboration because it is where governed and ungoverned practice diverge over time. As organizations report that large fractions of new code are AI-generated, the volume of code entering maintenance grows rapidly, making maintainability more important, not less \cite{borg2026echoes}. Under the spec-anchored invariant, drift between intent and implementation is detected at edit time rather than discovered archaeologically; under spec-as-source, maintenance becomes specification change plus regeneration, keeping the implementation permanently synchronized with design intent \cite{piskala2026spec}. Regeneration bounded by $\val$ is, in effect, continuous refactoring at low marginal human cost (DP5). The counter-evidence is equally instructive: the null maintainability result of \cite{borg2026echoes} was obtained under conventional review discipline, which is itself a (human-enforced) governance regime; the finding is therefore consistent with the SGRM's central claim that governance, not the presence of AI, is the moderating variable---while underscoring that the claim's causal form has not yet been isolated experimentally.

\subsection{Weighting the Evidence}
\label{sec:weighting}

The evidence supporting the design propositions is uneven in maturity, and a credible evaluation must say so. Three tiers can be distinguished. \emph{Replicated experimental findings}: the productivity effect of AI assistance \cite{peng2023impact,borg2026echoes} and the security risk of unreviewed generation \cite{pearce2022asleep,perry2023users,sandoval2023lost,fu2025security} rest on multiple independent designs and can be treated as established. \emph{Single rigorous studies}: the correctness gap \cite{liu2023your}, the practitioner-perceived unreliability of vibe coding \cite{fawzy2026vibe}, and the null maintainability result \cite{borg2026echoes} each rest on one strong design awaiting replication. \emph{Unreplicated case evidence}: the most striking pro-SDD quantifications---the 73\% security-defect reduction \cite{marri2026constitutional} and the one-person-squad multipliers \cite{vilasboas2026one}---derive from single case studies on preprint servers; they are verifiable as sources but not yet replicated as findings, and the argument of this article does not rest its weight on them. What the argument does rest on is the conjunction of the replicated tiers: generation is fast and fallible in specific, documented ways (F1--F4), and the SGRM mechanisms are targeted interdictions of exactly those documented ways.

\subsection{Boundary Conditions}
\label{sec:boundary}

The evaluation supports a conditional thesis, and the conditions matter. Specification governance carries authoring and maintenance costs of its own; Piskala's decision framework exists precisely because simpler approaches sometimes suffice \cite{piskala2026spec}. Table~\ref{tab:decision} summarizes the adoption decision surface implied by the analysis.

\begin{table}[htbp]
\centering
\small
\caption{Boundary conditions: when specification governance is and is not warranted.}
\label{tab:decision}
\begin{tabular}{L{4.7cm}L{4.0cm}L{3.6cm}}
\toprule
\rowcolor{rowtint}
\textbf{Context} & \textbf{Dominant need} & \textbf{Indicated practice} \\
\midrule
Ideation, prototyping, learning, creative exploration, MVP assembly & Speed, accessibility, low ceremony & Vibe coding / conversational generation \cite{fawzy2026vibe,meske2025vibe} \\
\addlinespace
Standalone utilities, internal scripts, short-lived tools & Adequacy under light use & Spec-first (lightweight specs, one-shot validation) \cite{piskala2026spec} \\
\addlinespace
Multi-developer products with sustained evolution & Reproducibility, maintainability & Spec-anchored (continuous validation) \cite{piskala2026spec} \\
\addlinespace
Enterprise, regulated, or safety-critical systems & Security, compliance, auditability, longevity & Spec-anchored to spec-as-source with constitutional constraints \cite{marri2026constitutional,belani2019requirements} \\
\bottomrule
\end{tabular}
\end{table}

\textbf{Answer to RQ3.} The published evidence supports specification-driven development as the foundation of AI-native \emph{enterprise} software engineering in the following precise sense: every enterprise quality characteristic of ISO/IEC~25010 faces a documented risk under ungoverned generation, and for every such risk the SGRM supplies a targeted mechanism grounded in at least one verified line of evidence, with the strongest quantifications (security-defect reduction, enterprise delivery outcomes) currently resting on unreplicated case studies. Outside the enterprise regime---ideation, prototyping, learning---the same evidence base affirms conversational generation as legitimate and valuable. The thesis is therefore conditional, and its condition is the presence of enterprise-grade quality-attribute demands.

% ================= DISCUSSION =================
\section{Discussion}
\label{sec:discussion}

\subsection{Implications for Practice}
\label{sec:practice}

For engineering organizations, the analysis reframes the adoption question. The choice is not ``AI or discipline'' but \emph{where to place the discipline}: upstream, in a machine-readable specification that governs generation, or downstream, in reactive review of an ever-growing volume of generated code. The evidence on reviewer overconfidence \cite{perry2023users} and on the sheer throughput of generation \cite{peng2023impact,borg2026echoes} suggests that downstream-only discipline does not scale: humans reviewing generated code at generation speed become the bottleneck and the weakest link simultaneously. The SGRM places the scarce resource---human judgment---at the two points where it has the highest leverage: authoring the contract (L1) and governing escalations (L4), while delegating the volume work of checking to the deterministic validator (L3). Practically, organizations can adopt the model incrementally along the rigor continuum of Figure~\ref{fig:continuum}: introducing specification templates and specification-derived tests (spec-first), then wiring continuous validation into integration pipelines (spec-anchored), and reserving spec-as-source for components where regeneration economics justify it \cite{piskala2026spec}. Tooling exists at every step: BDD frameworks and API-contract ecosystems, AI-assisted toolkits such as GitHub Spec Kit \cite{piskala2026spec}, constitutional constraint engines \cite{marri2026constitutional}, and multi-agent delivery pipelines \cite{vilasboas2026one}.

\subsection{Implications for the Competency Model of the Enterprise Engineer}
\label{sec:competency}

As automation assumes the burden of syntax production, professional competencies shift upstream. The AI-native era does not diminish the need for human expertise; it raises the bar for high-level technical judgment, system design, and verification \cite{alenezi2026rise,alenezi2026rethinking}. A systematic review of the field's trajectory documents the reorganization of practice, education, and workforce demands around orchestration, verification, and structured human--AI collaboration \cite{alenezi2026rise}, with the pedagogical objective shifting from producing code to critically evaluating, integrating, and governing AI-generated artifacts \cite{alenezi2026rethinking}. The intent-mediation analysis reaches a convergent conclusion: expertise reconstitutes as adaptive collaboration, valuing problem framing, output validation, and design thinking---the expert becomes an orchestrator who steers co-creation rather than a producer of syntax \cite{meske2025vibe}. The enterprise evidence adds a distributional warning: because AI assistants multiply the throughput of engineers who can specify, steer, and verify, while offering less to those who cannot yet perform oversight, expertise in specification authorship and verification becomes the scarce resource organizations must identify and cultivate \cite{vilasboas2026one,alenezi2026rethinking}. In SGRM terms, the engineer's defining functions concentrate at L1 and L4: articulating $(F,Q,K,\Sigma)$ precisely, and exercising accountable judgment when the loop escalates. Specification authorship---long treated as documentation overhead---becomes the discipline's core craft, and curricula that still equate programming competence with syntax fluency are training for the layer that has been automated.

\subsection{Open Questions and Research Agenda}
\label{sec:agenda}

The analysis exposes where the argument is thinnest, and each thin point is a research opportunity.

\begin{itemize}
\item \emph{Isolating the moderating variable.} The reconciliation of the debt evidence \cite{fawzy2026vibe} with the null maintainability result \cite{borg2026echoes}---that governance, not AI, moderates quality outcomes---is a hypothesis consistent with both datasets, not a tested finding. Longitudinal and controlled studies that manipulate review and specification discipline directly are the single most valuable next experiment.
\item \emph{Replication of SDD case results.} The 73\% security-defect reduction \cite{marri2026constitutional} and the one-person-squad outcomes \cite{vilasboas2026one} require replication across organizations, domains, and teams before they can carry prescriptive weight.
\item \emph{Specification languages.} The model presupposes specifications that are simultaneously human-authorable, machine-verifiable, and LLM-legible; designing such languages---and measuring the authoring cost they impose---is an open design problem descended from the expressiveness--tractability trade-off of program synthesis \cite{gulwani2017program,solarlezama2008program}.
\item \emph{Verification-native generators.} Current models are trained overwhelmingly on syntax; semantically grounded models that can participate in their own verification would shift work from L3 into L2 \cite{kessel2025morescient}, and N-version schemes suggest how generation diversity itself can be a verification resource \cite{kessel2024nversion}.
\item \emph{Skill formation and cognitive debt.} If routine implementation is delegated, how do engineers acquire the judgment the model concentrates at L1 and L4? The cognitive-debt risk flagged experimentally \cite{borg2026echoes} and the pedagogy of specification authorship \cite{alenezi2026rethinking} are unresolved.
\item \emph{Accountability.} Governance of the tooling layer itself---who is accountable when agents generate, modify, or recommend code---is an emerging research area currently mediated through providers' terms of service \cite{alenezi2026rethinking}.
\end{itemize}

Beyond these, the deepest open question is economic: the SGRM relocates cost from implementation to specification, and the discipline lacks measurement instruments for specification effort comparable to those it has for code. Brooks's essence--accident distinction predicts this is where the irreducible cost will pool \cite{brooks1987no}; quantifying it is prerequisite to honest cost--benefit analysis of the tiers in Table~\ref{tab:decision}.

% ================= THREATS =================
\section{Threats to Validity}
\label{sec:threats}

\textbf{Construct validity.} The central constructs---``vibe coding,'' ``specification governance,'' ``enterprise-grade''---are young and contested. The article mitigates this by anchoring each to published definitions (\cite{meske2025vibe,fawzy2026vibe} for vibe coding; \cite{piskala2026spec} for the rigor tiers; ISO/IEC~25010 \cite{iso25010} for quality attributes) and by formalizing specification governance (Definitions~\ref{def:spec}--\ref{def:governance}) so that the evaluation cannot equivocate between practices. Residual risk remains that practitioner usage of these terms will drift from the definitions adopted here.

\textbf{Internal validity.} The evaluation is analytical: it argues from published findings to the adequacy of designed mechanisms, in the informed-argument mode of design science \cite{hevner2004design}. No causal claim in this article has been tested by an experiment conducted for this article. The sharpest internal threat is the moderating-variable inference of Section~\ref{sec:counterevidence}: the proposition that governance, not AI assistance, moderates quality outcomes is consistent with the evidence but not isolated by it. The article marks this and the other unreplicated links explicitly (Section~\ref{sec:weighting}) rather than presenting them as established.

\textbf{External validity.} The empirical corpus over-represents certain settings: controlled experiments with professional developers on bounded tasks \cite{peng2023impact,borg2026echoes}, security benchmarks on scenario suites \cite{pearce2022asleep}, one regulated financial enterprise \cite{vilasboas2026one}, and one banking-microservices case \cite{marri2026constitutional}. Generalization to other domains, organization sizes, and regulatory regimes is plausible but unestablished. The grey-literature evidence \cite{fawzy2026vibe}, while systematically collected, inherits the selection biases of practitioner self-report \cite{garousi2019guidelines}.

\textbf{Reliability of the corpus.} The field moves quickly, and several load-bearing sources are preprints under review. Two disciplines limit the exposure: every reference was verified against a primary source before inclusion, with unverifiable claims excluded outright; and the evidence-maturity weighting of Section~\ref{sec:weighting} prevents unreplicated results from carrying the argument. Nonetheless, conclusions drawn from single case studies must be expected to be revised as replications arrive, and the reference model is deliberately stated so that such revisions bear on the strength, not the structure, of its propositions.

% ================= CONCLUSION =================
\section{Conclusion}
\label{sec:conclusion}

The future of software engineering is unlikely to abandon engineering in favor of unrestricted natural-language programming, and equally unlikely to retreat from AI-mediated generation. This article has argued, on the basis of a systematically verified evidence corpus, that the resolution of this tension is neither ``code-first'' nor ``prompt-first'' but \emph{specification-first}. The verified record identifies four recurring failure modes of ungoverned conversational generation---a productivity--reliability paradox, architectural erosion, quantified security exposure, and compounding technical debt---sharing a single causal structure: a stochastic generator whose acceptance criterion, observed behavior, is strictly weaker than the quality attributes the enterprise requires (RQ1). The Specification Governance Reference Model consolidates the remedy into a formal, tool-independent architecture: specifications as four-component contracts, a stochastic generator enclosed by a deterministic validation boundary, three progressively stronger rigor tiers, a closed regeneration loop that degrades safely to human governance, and traceability as a byproduct of process (RQ2). Evaluated analytically against ISO/IEC~25010, every enterprise quality characteristic maps to a documented risk of ungoverned generation, a targeted SGRM mechanism, and at least one verified line of evidence---with replicated experimental findings carrying the argument's weight and unreplicated case results (a 73\% security-defect reduction under constitutional constraints; halved time-to-market under spec-governed agentic delivery) marking its promising but provisional frontier (RQ3).

The thesis is deliberately conditional. Vibe coding is a genuine socio-technical innovation---for ideation, prototyping, learning, and the democratization of software creation, the evidence affirms it. But where testability, scalability, security, maintainability, traceability, and regulatory compliance are required, specification governance is the mechanism by which the stochastic fluency of generative models is bound to deterministic, auditable engineering. In that regime, the specification becomes the primary engineering artifact, AI agents automate implementation within its guardrails, and the engineer's defining work migrates to articulating contracts and exercising accountable judgment. The oldest insight of the discipline survives its newest technology intact: software quality is designed and demanded, never merely emitted.

% ================= REFERENCES =================
\bibliographystyle{unsrt}
\bibliography{references}

@article{alenezi2026rise,
  author  = {Alenezi, Mamdouh},
  title   = {The Rise of {AI}-Native Software Engineering: Implications for Practice, Education, and the Future Workforce},
  journal = {arXiv preprint arXiv:2606.12986},
  year    = {2026},
  url     = {https://arxiv.org/abs/2606.12986}
}

@article{alenezi2026rethinking,
  author  = {Alenezi, Mamdouh},
  title   = {Rethinking Software Engineering for Agentic {AI} Systems},
  journal = {arXiv preprint arXiv:2604.10599},
  year    = {2026},
  url     = {https://arxiv.org/abs/2604.10599}
}

@inproceedings{belani2019requirements,
  author    = {Belani, Hrvoje and Vukovic, Marin and Car, {\v{Z}}eljka},
  title     = {Requirements Engineering Challenges in Building {AI}-Based Complex Systems},
  booktitle = {Proceedings of the 27th IEEE International Requirements Engineering Conference Workshops (REW)},
  pages     = {252--255},
  publisher = {IEEE},
  year      = {2019}
}

@inproceedings{bhargavan2016formal,
  author    = {Bhargavan, Karthikeyan and Delignat-Lavaud, Antoine and Fournet, C{\'e}dric and Gollamudi, Anitha and Gonthier, Georges and Kobeissi, Nadim and Kulatova, Natalia and Rastogi, Aseem and Sibut-Pinote, Thomas and Swamy, Nikhil and Zanella-B{\'e}guelin, Santiago},
  title     = {Formal Verification of Smart Contracts: Short Paper},
  booktitle = {Proceedings of the 2016 ACM Workshop on Programming Languages and Analysis for Security (PLAS '16)},
  pages     = {91--96},
  publisher = {ACM},
  year      = {2016}
}

@book{boehm1981software,
  author    = {Boehm, Barry W.},
  title     = {Software Engineering Economics},
  publisher = {Prentice-Hall},
  address   = {Englewood Cliffs, NJ},
  year      = {1981}
}

@article{borg2026echoes,
  author  = {Borg, Markus and Hewett, Dave and Hagatulah, Nadim and Couderc, No{\'e}mie and S{\"o}derberg, Emma and Graham, Donald and Kini, Uttam and Farley, Dave},
  title   = {Echoes of {AI}: Investigating the Downstream Effects of {AI} Assistants on Software Maintainability},
  journal = {Empirical Software Engineering},
  volume  = {31},
  number  = {6},
  year    = {2026},
  note    = {Preprint: arXiv:2507.00788}
}

@article{brooks1987no,
  author  = {Brooks, Frederick P.},
  title   = {No Silver Bullet: Essence and Accidents of Software Engineering},
  journal = {Computer},
  volume  = {20},
  number  = {4},
  pages   = {10--19},
  year    = {1987}
}

@article{bucchiarone2020grand,
  author  = {Bucchiarone, Antonio and Cabot, Jordi and Paige, Richard F. and Pierantonio, Alfonso},
  title   = {Grand Challenges in Model-Driven Engineering: An Analysis of the State of the Research},
  journal = {Software and Systems Modeling},
  volume  = {19},
  number  = {1},
  pages   = {5--13},
  year    = {2020}
}

@article{chen2021evaluating,
  author  = {Chen, Mark and Tworek, Jerry and Jun, Heewoo and Yuan, Qiming and Pinto, Henrique Ponde de Oliveira and Kaplan, Jared and others},
  title   = {Evaluating Large Language Models Trained on Code},
  journal = {arXiv preprint arXiv:2107.03374},
  year    = {2021}
}

@inproceedings{clelandhuang2014software,
  author    = {Cleland-Huang, Jane and Gotel, Orlena C. Z. and Huffman Hayes, Jane and M{\"a}der, Patrick and Zisman, Andrea},
  title     = {Software Traceability: Trends and Future Directions},
  booktitle = {Future of Software Engineering Proceedings (FOSE 2014, ICSE)},
  pages     = {55--69},
  publisher = {ACM},
  year      = {2014}
}

@article{cotroneo2025human,
  author  = {Cotroneo, Domenico and Improta, Cristina and Liguori, Pietro},
  title   = {Human-Written vs. {AI}-Generated Code: A Large-Scale Study of Defects, Vulnerabilities, and Complexity},
  journal = {arXiv preprint arXiv:2508.21634},
  year    = {2025}
}

@inproceedings{fan2023large,
  author    = {Fan, Angela and Gokkaya, Beliz and Harman, Mark and Lyubarskiy, Mitya and Sengupta, Shubho and Yoo, Shin and Zhang, Jie M.},
  title     = {Large Language Models for Software Engineering: Survey and Open Problems},
  booktitle = {Proceedings of the IEEE/ACM International Conference on Software Engineering: Future of Software Engineering (ICSE-FoSE)},
  pages     = {31--53},
  publisher = {IEEE},
  year      = {2023}
}

@inproceedings{fawzy2026vibe,
  author    = {Fawzy, Ahmed and Tahir, Amjed and Blincoe, Kelly},
  title     = {Vibe Coding in Practice: Motivations, Challenges, and a Future Outlook---A Grey Literature Review},
  booktitle = {Proceedings of the 48th International Conference on Software Engineering: Software Engineering in Practice (ICSE-SEIP '26)},
  publisher = {ACM},
  year      = {2026},
  note      = {Preprint: arXiv:2510.00328}
}

@article{fu2025security,
  author  = {Fu, Yujia and Liang, Peng and Tahir, Amjed and Li, Zengyang and Shahin, Mojtaba and Yu, Jiaxin and Chen, Jinfu},
  title   = {Security Weaknesses of Copilot-Generated Code in {GitHub} Projects: An Empirical Study},
  journal = {ACM Transactions on Software Engineering and Methodology},
  year    = {2025},
  publisher = {ACM}
}

@article{garousi2019guidelines,
  author  = {Garousi, Vahid and Felderer, Michael and M{\"a}ntyl{\"a}, Mika V.},
  title   = {Guidelines for Including Grey Literature and Conducting Multivocal Literature Reviews in Software Engineering},
  journal = {Information and Software Technology},
  volume  = {106},
  pages   = {101--121},
  year    = {2019}
}

@article{gulwani2017program,
  author  = {Gulwani, Sumit and Polozov, Oleksandr and Singh, Rishabh},
  title   = {Program Synthesis},
  journal = {Foundations and Trends in Programming Languages},
  volume  = {4},
  number  = {1--2},
  pages   = {1--119},
  year    = {2017}
}

@article{hassan2024towards,
  author  = {Hassan, Ahmed E. and Lin, Dayi and Rajbahadur, Gopi Krishnan and Gallaba, Keheliya and C{\^o}go, Filipe R. and Chen, Boyuan and others},
  title   = {Towards {AI}-Native Software Engineering ({SE} 3.0): A Vision and a Challenge Roadmap},
  journal = {arXiv preprint arXiv:2410.06107},
  year    = {2024}
}

@article{hevner2004design,
  author  = {Hevner, Alan R. and March, Salvatore T. and Park, Jinsoo and Ram, Sudha},
  title   = {Design Science in Information Systems Research},
  journal = {MIS Quarterly},
  volume  = {28},
  number  = {1},
  pages   = {75--105},
  year    = {2004}
}

@article{hou2024large,
  author  = {Hou, Xinyi and Zhao, Yanjie and Liu, Yue and Yang, Zhou and Wang, Kailong and Li, Li and Luo, Xiapu and Lo, David and Grundy, John and Wang, Haoyu},
  title   = {Large Language Models for Software Engineering: A Systematic Literature Review},
  journal = {ACM Transactions on Software Engineering and Methodology},
  volume  = {33},
  number  = {8},
  pages   = {220:1--220:79},
  year    = {2024}
}

@misc{iso25010,
  author       = {{ISO/IEC}},
  title        = {{ISO/IEC 25010:2023} --- Systems and Software Engineering --- Systems and Software Quality Requirements and Evaluation ({SQuaRE}) --- Product Quality Model},
  howpublished = {International Organization for Standardization, Geneva},
  year         = {2023}
}

@inproceedings{karampatsis2020big,
  author    = {Karampatsis, Rafael-Michael and Babii, Hlib and Robbes, Romain and Sutton, Charles and Janes, Andrea},
  title     = {Big Code != Big Vocabulary: Open-Vocabulary Models for Source Code},
  booktitle = {Proceedings of the ACM/IEEE 42nd International Conference on Software Engineering (ICSE '20)},
  pages     = {1073--1085},
  publisher = {ACM},
  year      = {2020}
}

@misc{karpathy2025vibe,
  author       = {Karpathy, Andrej},
  title        = {There's a new kind of coding I call ``vibe coding''},
  howpublished = {X (formerly Twitter) post, \url{https://x.com/karpathy/status/1886192184808149383}},
  month        = feb,
  year         = {2025}
}

@article{kessel2024nversion,
  author  = {Kessel, Marcus and Atkinson, Colin},
  title   = {N-Version Assessment and Enhancement of Generative {AI}},
  journal = {IEEE Software},
  volume  = {42},
  number  = {2},
  pages   = {76--83},
  year    = {2025},
  note    = {Preprint: arXiv:2409.14071}
}

@article{kessel2025morescient,
  author  = {Kessel, Marcus and Atkinson, Colin},
  title   = {Morescient {GAI} for Software Engineering},
  journal = {ACM Transactions on Software Engineering and Methodology},
  year    = {2025},
  note    = {DOI: 10.1145/3709354}
}

@techreport{kitchenham2007guidelines,
  author      = {Kitchenham, Barbara and Charters, Stuart},
  title       = {Guidelines for Performing Systematic Literature Reviews in Software Engineering},
  institution = {Keele University and Durham University},
  number      = {EBSE-2007-01},
  year        = {2007}
}

@inproceedings{lemieux2023codamosa,
  author    = {Lemieux, Caroline and Inala, Jeevana Priya and Lahiri, Shuvendu K. and Sen, Siddhartha},
  title     = {{CodaMosa}: Escaping Coverage Plateaus in Test Generation with Pre-Trained Large Language Models},
  booktitle = {Proceedings of the 45th IEEE/ACM International Conference on Software Engineering (ICSE '23)},
  pages     = {919--931},
  publisher = {IEEE},
  year      = {2023}
}

@article{li2022competition,
  author  = {Li, Yujia and Choi, David and Chung, Junyoung and Kushman, Nate and Schrittwieser, Julian and Leblond, R{\'e}mi and others},
  title   = {Competition-Level Code Generation with {AlphaCode}},
  journal = {Science},
  volume  = {378},
  number  = {6624},
  pages   = {1092--1097},
  year    = {2022}
}

@inproceedings{liu2023your,
  author    = {Liu, Jiawei and Xia, Chunqiu Steven and Wang, Yuyao and Zhang, Lingming},
  title     = {Is Your Code Generated by {ChatGPT} Really Correct? Rigorous Evaluation of Large Language Models for Code Generation},
  booktitle = {Advances in Neural Information Processing Systems 36 (NeurIPS 2023)},
  year      = {2023}
}

@article{lukasczyk2023empirical,
  author  = {Lukasczyk, Stephan and Kroi{\ss}, Florian and Fraser, Gordon},
  title   = {An Empirical Study of Automated Unit Test Generation for {Python}},
  journal = {Empirical Software Engineering},
  volume  = {28},
  number  = {2},
  pages   = {36},
  year    = {2023}
}

@article{marri2026constitutional,
  author  = {Marri, Sudheer Reddy},
  title   = {Constitutional Spec-Driven Development: Enforcing Security by Construction in {AI}-Assisted Code Generation},
  journal = {arXiv preprint arXiv:2602.02584},
  year    = {2026},
  url     = {https://arxiv.org/abs/2602.02584}
}

@article{meske2025vibe,
  author  = {Meske, Christian and Hermanns, Tobias and von der Weiden, Esther and Loser, Kai-Uwe and Berger, Thorsten},
  title   = {Vibe Coding as a Reconfiguration of Intent Mediation in Software Development: Definition, Implications, and Research Agenda},
  journal = {IEEE Access},
  volume  = {13},
  pages   = {213242--213259},
  year    = {2025},
  note    = {DOI: 10.1109/ACCESS.2025.3645466}
}

@article{meyer1992applying,
  author  = {Meyer, Bertrand},
  title   = {Applying ``Design by Contract''},
  journal = {Computer},
  volume  = {25},
  number  = {10},
  pages   = {40--51},
  year    = {1992}
}

@article{ozkaya2023application,
  author  = {Ozkaya, Ipek},
  title   = {Application of Large Language Models to Software Engineering Tasks: Opportunities, Risks, and Implications},
  journal = {IEEE Software},
  volume  = {40},
  number  = {3},
  pages   = {4--8},
  year    = {2023}
}

@article{parnas1972criteria,
  author  = {Parnas, David L.},
  title   = {On the Criteria To Be Used in Decomposing Systems into Modules},
  journal = {Communications of the ACM},
  volume  = {15},
  number  = {12},
  pages   = {1053--1058},
  year    = {1972}
}

@inproceedings{pearce2022asleep,
  author    = {Pearce, Hammond and Ahmad, Baleegh and Tan, Benjamin and Dolan-Gavitt, Brendan and Karri, Ramesh},
  title     = {Asleep at the Keyboard? Assessing the Security of {GitHub} {Copilot}'s Code Contributions},
  booktitle = {Proceedings of the 2022 IEEE Symposium on Security and Privacy (SP)},
  pages     = {754--768},
  publisher = {IEEE},
  year      = {2022}
}

@article{peng2023impact,
  author  = {Peng, Sida and Kalliamvakou, Eirini and Cihon, Peter and Demirer, Mert},
  title   = {The Impact of {AI} on Developer Productivity: Evidence from {GitHub} {Copilot}},
  journal = {arXiv preprint arXiv:2302.06590},
  year    = {2023}
}

@inproceedings{perry2023users,
  author    = {Perry, Neil and Srivastava, Megha and Kumar, Deepak and Boneh, Dan},
  title     = {Do Users Write More Insecure Code with {AI} Assistants?},
  booktitle = {Proceedings of the 2023 ACM SIGSAC Conference on Computer and Communications Security (CCS '23)},
  pages     = {2785--2799},
  publisher = {ACM},
  year      = {2023}
}

@article{piskala2026spec,
  author  = {Piskala, Deepak Babu},
  title   = {Spec-Driven Development: From Code to Contract in the Age of {AI} Coding Assistants},
  journal = {arXiv preprint arXiv:2602.00180},
  year    = {2026},
  url     = {https://arxiv.org/abs/2602.00180}
}

@inproceedings{sandoval2023lost,
  author    = {Sandoval, Gustavo and Pearce, Hammond and Nys, Teo and Karri, Ramesh and Garg, Siddharth and Dolan-Gavitt, Brendan},
  title     = {Lost at {C}: A User Study on the Security Implications of Large Language Model Code Assistants},
  booktitle = {Proceedings of the 32nd USENIX Security Symposium},
  pages     = {2205--2222},
  publisher = {USENIX Association},
  year      = {2023}
}

@inproceedings{sculley2015hidden,
  author    = {Sculley, D. and Holt, Gary and Golovin, Daniel and Davydov, Eugene and Phillips, Todd and Ebner, Dietmar and Chaudhary, Vinay and Young, Michael and Crespo, Jean-Fran{\c{c}}ois and Dennison, Dan},
  title     = {Hidden Technical Debt in Machine Learning Systems},
  booktitle = {Advances in Neural Information Processing Systems 28 (NIPS 2015)},
  pages     = {2503--2511},
  year      = {2015}
}

@phdthesis{solarlezama2008program,
  author = {Solar-Lezama, Armando},
  title  = {Program Synthesis by Sketching},
  school = {University of California, Berkeley},
  year   = {2008}
}

@article{vilasboas2026one,
  author  = {Vilas Boas, Marcelo and Pinto, Gustavo and Monteiro, Edward Roberto and Carida, Vinicius Fernandes and Ribeiro, Danilo},
  title   = {One Developer Is All You Need: A Case Study of an {AI}-Augmented One-Person Squad in a Brownfield Enterprise},
  journal = {arXiv preprint arXiv:2605.18461},
  year    = {2026},
  url     = {https://arxiv.org/abs/2605.18461}
}

@book{wieringa2014design,
  author    = {Wieringa, Roel J.},
  title     = {Design Science Methodology for Information Systems and Software Engineering},
  publisher = {Springer},
  address   = {Berlin, Heidelberg},
  year      = {2014}
}

@inproceedings{zan2023large,
  author    = {Zan, Daoguang and Chen, Bei and Zhang, Fengji and Lu, Dianjie and Wu, Bingchao and Guan, Bei and Wang, Yongji and Lou, Jian-Guang},
  title     = {Large Language Models Meet {NL2Code}: A Survey},
  booktitle = {Proceedings of the 61st Annual Meeting of the Association for Computational Linguistics (ACL 2023)},
  pages     = {7443--7464},
  publisher = {ACL},
  year      = {2023}
}

\end{document}